# Gentle patterning approaches towards compatibility with bio-organic materials and their environmental aspects

*Artem K. Grebenko*, Konstantin A. Motovilov, Anton V. Bubis, Albert G. Nasibulin***

Dr. A. K. Grebenko, A. V. Bubis, Prof. A. G. Nasibulin
Skolkovo Institute of Science and Technology, Nobel str. 3, 121205, Moscow, Russia
E-mail: *artem.grebenko@skoltech.ru, **a.nasibulin@skoltech.ru

Dr. A. K. Grebenko, Dr. K. A. Motovilov
Center for Photonics and 2D Materials, Moscow Institute of Physics and Technology, Institute Lane 9, 141700, Dolgoprudny, Russia

A. V. Bubis
Institute of Solid State Physics, Russian Academy of Sciences, 2 Academician Ossipyan str., 142432, Chernogolovka, Russia

Prof. A. G. Nasibulin
Aalto University, P.O. Box 16100, FI-00076 Aalto, Finland

Advances in material science, bio-electronic, and implantable medicine combined with recent requests for eco-friendly materials and technologies inevitably formulate new challenges for nano- and micro-patterning techniques. Overall, the importance of creating micro and nanostructures is motivated by a large manifold of fundamental and applied properties accessible only at the nanoscale. Lithography is a crucial family of fabrication methods to create prototypes and produce devices on an industrial scale. The pure trend in the miniaturization of critical electronic semiconducting components is recently enhanced by implementing bio-organic systems in electronics. So far, significant efforts have been made to find novel lithographic approaches and develop old ones to reach compatibility with delicate bio-organic systems and minimize the impact on the environment. Herein, we briefly review such delicate materials and sophisticated patterning techniques.

## 1. Introduction

Among 14 grand challenges delineated by the National Academy of Engineering (USA) for the 21st century,[1] at least four of them - advanced health informatics, reverse-engineering of the brain, enhancement of the virtual reality, and advancement of personalized learning - require improvement of the methods to transmit and process information generated by a human organism. Indeed, the recent decade has brought rapid development of various "electroceuticals" far beyond familiar heart pacemakers and cochlear implants.[2,3] Deep brain, spinal cord, sacral and vagus nerves, and transcutaneous electrical nerve stimulators are now applied for the treatment of chronic pain,[4] restoration of sensory and motor function,[5] treatment of Parkinson's disease,[6] various urinary symptoms,[7] inflammatory disorders of joints and gastrointestinal tract, migraine, and other headaches.[8] USA Food and Drug Administration (FDA) approves new electroceutical systems from market leaders and startup companies every month.[9] Great research efforts are made to effectively implement optimized stimuli waveforms[10] and frequency modulation.[11]

The development of flexible and stretchable bio-electronic components and devices is carried out in two magistral directions. The first one relies on intrinsically deformable bio-organic materials.[12–18] The second pathway implicates inorganic materials that are naturally non-stretchable, like silicon, transferred into a flexible/stretchable substrate[19–21] and nanostructures, such as single-walled carbon nanotubes, placed on top of stretchable polymeric substrates.[22,23] In the case of silicon-based solutions, solid islands are connected by bridges in the form of serpentine, fractals, and waves. These island-bridge structures become relatively flexible and stretchable, retaining the main advantages of inorganics fabricated by matured technologies: high electrical conductivity, stability, and slow aging of the components. From the other point of view, low biocompatibility and poor degradability are the flip side effects of the inorganics. As usually expected, the most successful result can be achieved by coupling advantages of both pathways.

Despite the significant development of electroceuticals, we should still overcome several challenges to achieve miniaturization and autonomy. The first one is low spatial resolution in electrical circuits of current devices. Due to high neural tissue density, naturally unconnected information paths are close in space. For instance, the vagus nerve consists of about 100000 fibers innervating many different organs.[2] Its electrical activation or inhibition leads to unclear clinical consequences. Deep brain stimulation techniques constantly meet this issue.[24] Side effects often include problems in cognition and emotional spheres since many other cells are affected by electrical stimuli besides those needed the treatment. Thus, high-density electrode arrays should consist of independent mesoscale soft and flexible conductors selectively interacting with individual neural fibers. Production of such arrays with the required spatial resolution, using classical lithography techniques and conventional materials, still needs to be developed in the nearest future.[25]

The second serious challenge is getting the ability to receive energy from natural physiological processes. According to a recent review of power sources for wearable and implantable biosensors,[26] the last significant achievement performed in vivo was reported ten years ago.[27] One of the recently found sideways to solve this complex problem came from triboelectrics.[28–32] The development of triboelectric nanogenerators (TENG) leaves hope to power low-energy bioelectronic devices employing the kinetic energy of muscles. However, a bioelectronic device located away from actively contracting striated muscle tissue, which powers TENG, requires long electronic wires.

Finally, there is a great need to change the frequency of electrical stimuli immediately after the change in the physiological status. It demands an integration of sensors, actuators, logical modules, and a power supply into a single device. At the moment, we are far from that, but some of the newly developed methods discussed in the current review give us hope to achieve this goal in the upcoming decade. Given the complexity and unique peculiarities of the human body, bioelectronics should become tightly adaptable as other medical therapies.

Current plans towards conjugating neuro- and nanoscience are based mainly on various materials, including plasmonic and upconverting nanoparticles, quantum dots, magnetic nanoparticles, etc. The great potential of different techniques has been recently articulated in a perspective article by Garcia-Etxarri and Yuste.[33] Undoubtedly, we are expecting the creation of high-precision sensors and actuators for neuronal cells with spatial and temporal resolutions unachievable earlier. However, all discussed techniques suffer from a strong dependence on external energy sources: alternating magnetic field, ultrasound, or visible/near-infrared radiation. Therefore, we should build autonomous multilayered logical schemes allowing complementary therapy without external intervention in a similar manner as suggested by theranostics.[34] Due to its flexibility, scalability, and automation, most likely lithography might become the critical method in scalability and transition from laboratory to the industrial-grade fabrication of the therapeutic agents.

Bioelectronics is not the only field where green lithography techniques with the high spatial resolution are required. According to the report "The Global E-waste Monitor 2020: Quantities, flows and the circular economy potential report of Green electronics"[35] prepared by leading international organizations in the field of electronics, in 2019, the World industry produced 53.6 megatons of electrical and electronic waste (e-waste) materials hazardous to the environment. This ecological problem heralds a growing need to shift low-energy electronics manufacturing to eco-friendly materials, i.e., to develop "green electronics."[35,36]

Here, we review modern patterning solutions to overcome the challenges mentioned above regarding compatibility with bio-organic materials and their environmental friendliness. We present a detailed comparison of existing solutions, focusing on radiation-based patterning, the most widespread techniques in scientific and industrial communities. Finally, we review a few other approaches, which do not employ electron beam and ultraviolet radiation in their principle of operation.

## 2. Materials for patterning

As mentioned in the introduction, to develop flexible and stretchable bio-electronic devices, we need to solve the issues related to low spatial resolution, non-volatility, and independent tunability. Unfortunately, the existing methods of metallic contact creation or mesa-structures etching fail to deal with most bio-organic samples and do not allow to apply them in real-world devices.

### *2.1. Organic semiconductors*

First, let us discuss conductive organic materials,[37,38] an essential part of the contemporary microelectronics industry. This class of organics is much broader than presented in end-user devices and developing at an enormous pace. More intriguing fact is that the scientific community claims these very substances play a defining role in bioelectronics.[14,18,39,40] Great efforts were made to design a patterning approach, which allows precise template formation for integrated device applications.[41] However, still there are no general methods compatible with all or at least most valuable materials facilitating μm and sub- μm size features. Here, we could mention a couple of solutions developed for certain organic materials. For instance, direct electron-beam writing allows to form micro- and nanostructures in a metal-organic framework - zinc-imidazolate.[42] Adopting commercial resists, such as ZEP520 combined with polyvinyl alcohol (PVA) [43] or CYTOP™ (dielectric used in organic field-effect transistors),[44,45] has been recently demonstrated for some particular organic semiconductors (N,N′-Ditridecylperylene-3,4,9,10-tetracarboxylic diimide and pentacene).

### *2.2. Biomaterials*

Biological objects are often practically incompatible with traditional lithographic methods to create contacts for charge transport measurements. Neither focused ion beam (FIB) nor common photo- and e-beam lithography ensures a proper degree of compatibility and absence of detrimental impact. Almost every day, new biological materials are investigated in biochemistry and biophysics. We could mention tubulin microtubules,[46] multiheme cytochromes of the respiratory chains,[47] chlorosome photonic antenna complexes[48], photosynthetic complexes,[49] and DNA.[50] Relatively new objects are pili, filaments, and other extracellular outgrowths[51–53] of electrogenic bacteria, common extracellular membranes of linear electron-conductive bacterial colonies,[54] artificial and natural supramolecular peptide complexes.[55–58]

Researchers have actively studied the electron transport properties of biological objects for already 20 years. Nevertheless, very often experimental measurements of the conductivity cannot be easily explained and requires additional studies. Indeed, gallium ions used during FIB-based[59,60] deposition could burn an organic matter re-precipitated in the surrounding medium.[61] There are no fewer doubts when contemporary commercial resists are applied for delicate materials. Usually, solvents and processing treatments employed in a patterning protocol are far beyond the comfort zone of the bio-organic systems. The patch-clamp technique, utilized to measure ionic currents and electrochemical potential of membrane proteins,[62] might destroy the integrity of the investigated object. Until recently, we have not had the opportunity to create high-quality metallic contacts on bio-organic objects on a nano- and even micro-scale to measure their electronic conductivity without fear of being destroyed. Further we are going to specify few especially important cases of biological nano-systems and motivate design of novel patterning techniques by outstanding properties and perspective of this biostructures integration in microelectronics.

### *2.2.1. Peptides*

Almost 30 years ago, Ghadiri et al. published the milestone paper regarding cyclic octapeptides.[63] This work was the beginning of active research of peptides for creating supramolecular semiconducting materials. Peptides have become a basis for a new branch of nanophotonics, revealing biomedical nano-therapy tools and implantable integrated optical biochips.[64–66] The diversity of the peptide-based material applications goes far beyond the mentioned areas.[67,68] Utilization of natural amino acids helped solve problems associated with rapid proteolytic degradation of peptide materials in the body.[69] In turn, oxidative alternation of the side chains of amino acids may yield additional remarkable features like enhanced proton conductivity similar to melanin.[56] Among supramolecular peptide phases, widespread 1D (tubes, fibers, sticks) and 2D (flakes and meshes) self-ordered morphologies are good candidates for patterning. However, these structures are sensitive to minimal changes in the environmental parameters. Peptide patterning is a highly demanded technique for defining nanostructures for studies of cell adhesion mechanisms[70,71] and bio- and THz sensors.[72]

### 2.2.2. Biological nanowires

Protein and protein-lipid nanowire structures,[73,74] in which electron transfer is observed over significant distances (in terms of biological dimensions), also lack the developed methodological base of patterning. Indeed, being a convenient natural nanostructure, these objects require a design of reliable electrical contacts, which might imply deposition of non-metallic delicate conductors with a nanometer precision. Despite numerous studies of charge transport in multiheme cytochrome nanowires of Shewanella oneidensis MR-1[51,59,60,75–78] and extracellular conductive nanofilaments of Geobacter sulfurreducens,[79-84] the reliability of the measurements and understanding of fundamental transport mechanisms remain unclear. Artificial peptide nanowires have been synthesized and proved to be capable of charge transfer,[55,85,86] yet lacking non-destructive methods for contact fabrication. We should also mention problems of micro- and nano-device fabrication of machines and motors employing biological nanowires.[87]

### *2.2.3. DNA*

DNA is one of the classic patterning materials with a biological origin.[88,89] Thanks to the diversity of controlled self-organization methods in abundance existing in nanotechnology, one can obtain a wide range of stable structures on a DNA basis. Due to the possibility of error-free synthesis of primary structures, DNA has practically become a basic bio-organic construction material for nano-devices with computer-assisted design software.[90] An increase in the nuclease resistance demonstrated for some types of DNA molecules[91] gives us hope to fabricate devices, functioning under physiological conditions for a sufficiently long time. The emergence of soft lithographic techniques that are not destructive to the supramolecular DNA organization will significantly broaden the range of their nanotechnological applications.

## 3. Patterning Approaches

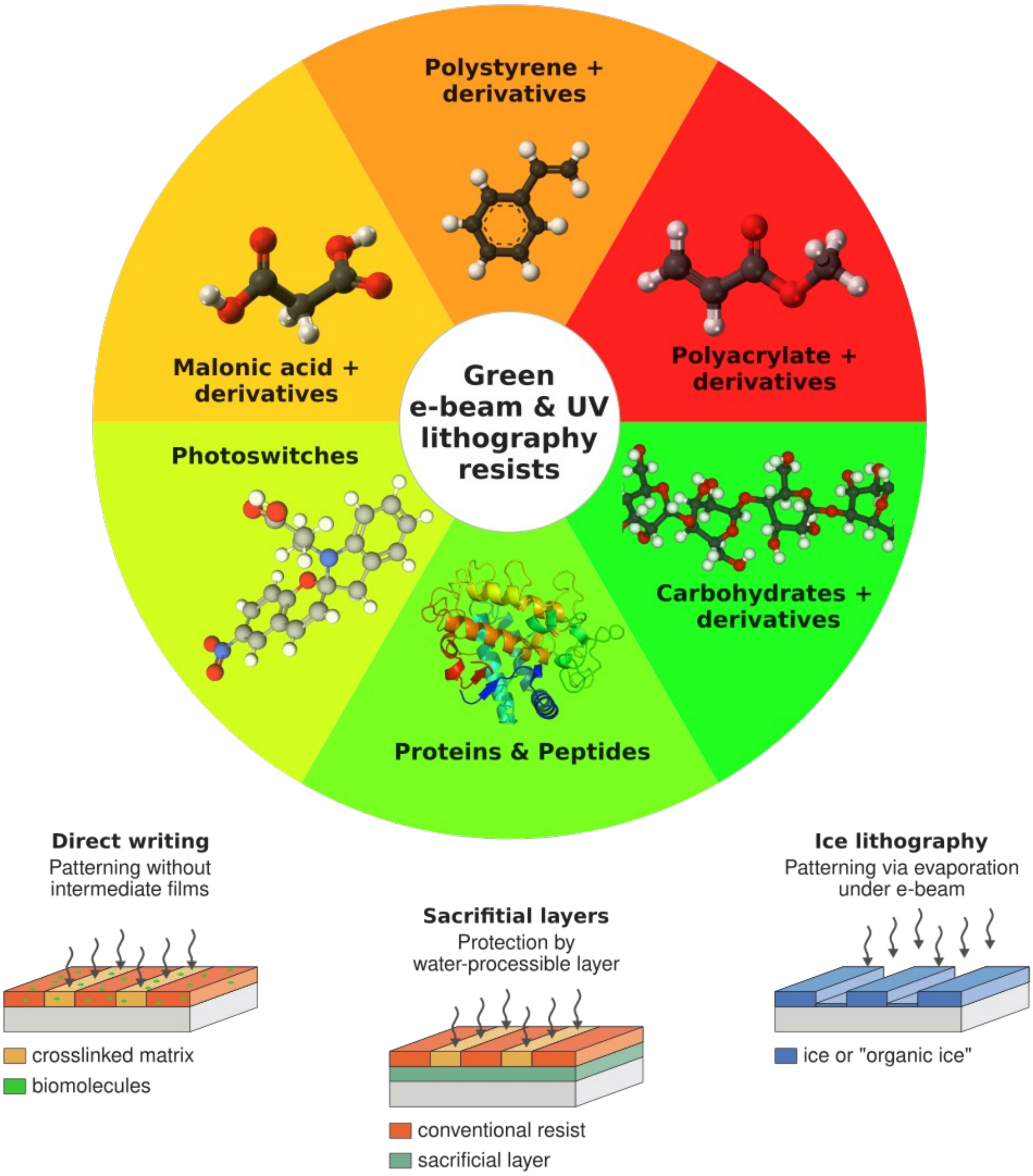

**Figure 1. Scheme of reviewed technological solutions towards green patterning approaches with enhanced compatibility to bio-organic materials.**

Radiation lithography[92] is the most widespread and reliable patterning method. Industrial-scale production, lab-scale prototyping, and research sample fabrication widely employ this method. By now, many fruitful solutions towards highly compatible green techniques have been designed (summarized graphically in **Figure 1**), including standard and unconventional approaches. The traditional way in the radiation lithography concept requires a particular intermediate film called resist, typically an organic one, utilized to define a mask on the substrate. We assigned colors from red to green to highlight the evolution of biocompatibility and eco-friendliness of the developed solutions regarding the compound forming a vital element of the technique – resist film (circular diagram in **Figure 1**). The solubility of the conventional resist depends on the molecular weight of the polymer, which changes during exposure. Several unconventional peculiar approaches are placed outside of the central diagram - they help overcome limitations of the standard techniques by a tricky workaround.

A resist can be of a negative tone when the radiation modifies exposed regions to withstand a subsequent dissolution process called development. In contrast, for a positive tone resist irradiated areas are dissolved during the development, while the pristine part of the film stays intact. A solvent used at the development stage is a developer. After the development, patterned resist film is used as a mask for consequent modification by either top-down approaches (e.g., plasma etching, ion implantation) or bottom-up techniques (e.g., deposition followed by lift-off). Finally, a solvent called remover entirely dissolves the resist film. We should also mention specific steps often taken before/after resist exposure (e.g., baking), denoted further as pre-exposure/post-exposure, and right after the development, denoted further as post-development. It is worth mentioning that the applicability of the patterning technique to a particular sample is limited not only by the chemicals but also by these additional technological steps. Graphically the radiation lithography technique is illustrated in **Figure 2**.

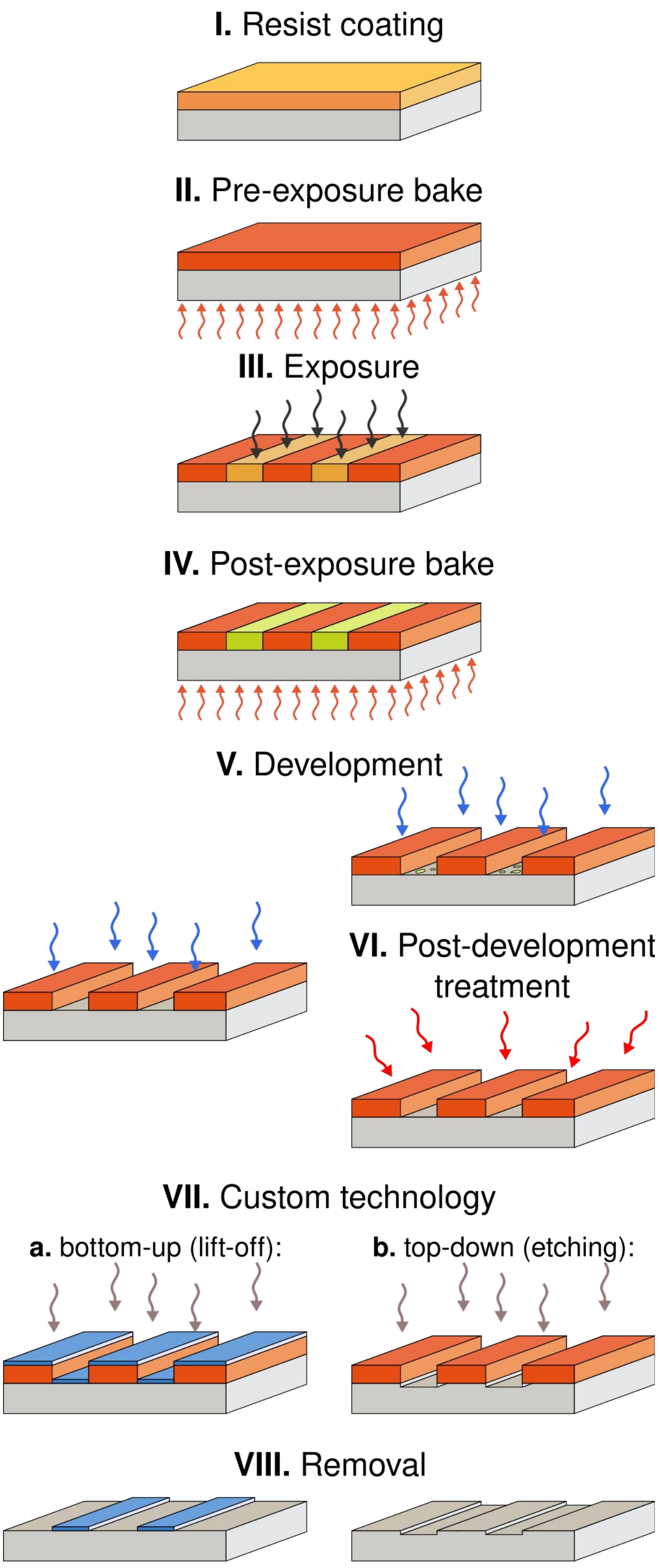

**Figure 2. Schematic illustration of the radiation lithography steps**. **(I)** First, the resist film is deposited on the surface of a target material by means of spin-coating, vapor deposition, Langmuir-Blodgett technique, or any other methods. **(II)** Next, most conventional approaches require the pre-exposure baking procedure to remove excess solvent or activate additives. **(III)** Further, the resist is exposed to corresponding radiation, resulting in the change of solubility properties. **(IV)** After that some methods require post-exposure baking for activation of additives. **(V)** The process is followed by the development, i.e. chemical or physical modification of the resist film. **(VI)** If after the development a significant amount of the resist residues is left on the surface, the post-

development treatment should be performed. **(VII)** At this stage subsequent fabrication routines through the developed resist film are performed: **(VIIa)** a bottom-up technique (e.g., lift-off), in which another material is deposited into opened windows, resulting in bulging structures after removing of the non-developed resist film **(VIIIa)**; **(VIIb)** the top-down technique, illustrated by the etching process, in which the substrate material is removed through opened windows, resulting in trench structures after eliminating the non-developed resist film **(VIIIb)**.

### 3.1. Optical and DUV/EUV lithography

Optical lithography[93] (OL), including deep and extreme ultraviolet[94] (DUV and EUV), employs light to modify the solubility of the resist. OL allows patterning with 0.5 µm resolution in direct writing setups and regular mask projection schemes. There are three principal approaches to prepare water-soluble photoresists, which, being photochemical reactions, usually follow the radical mechanism: photoactivated cross-linking,[95–98] photo-elimination/photo-deprotection,[99–101] and photo-induced rearrangement/photoisomerization.[102]

Aqueous-based development always requires a change in the resist polarity or phase state. In turn, biocompatibility demands polarity change to be soft, without significant pH shift and formation of excited radicals. That significantly limits the utilization of the state-of-the-art chemically amplified resists, since employed acidic anhydrides[103] might react with nucleophilic groups on the surface of proteins and other biomaterials. On the other hand, acylation reactions are entirely physiological, which can be selectively reversed using appropriate enzymes.

Studies carried out during the last two decades revealed some optimism in biocompatibility of photoswitching compounds[104,105] . Kinetic measurements combined with photokinetic simulations showed that interference lithography using conventional light sources (e.g., a green laser) on a spirothiopyran maleimide-based system delivered high-resolution features (~45 nm lines) in thick films (tens of µm) over large areas (hundreds of µm).[106] UV-Vis irradiation of some azobenzene or spiropyran-based systems induced phase transitions from solid to liquid even at temperatures close to physiological ones.[107–113]

We should also note that photoisomerization of the spiropyran system brings the dipole moment change $\Delta\mu$ = 8-15 D,[104] which makes it perfect candidate for the aqueous-based photoresist. Biocompatible hydrogels often employ this feature to control conformation and activity of biomolecules, but they have never  been yet employed as a resist. Another peculiar strategy is based on so-called light-stabilized dynamic materials, in which a cleavage of solid covalently cross-linked material into viscous liquid occurs upon switching the irradiation off.[114]

In our opinion, the potential of photo-switchable chemical groups to create water-soluble photoresists remains underestimated. One of the main tasks of bioelectronics is to master the fabrication of devices directly in the body environment or on the surgical operation site. In this regard, the combination of photoresist and hydrogel properties of photo-switches are seemed to be promising.

Further, we survey several historically significant solutions towards water-processable resists and start with papers by Wilson's group from the University of Texas at Austin (USA).[115,116] They incorporated a photoactive catalyst such as a photoacid generator into a resist film. As an example, Wilson's group designed an environmentally-friendly negative tone resist[115] that consisted of three water-soluble components: a polymer (poly(methyl acrylamidoglycolate methyl ether), [poly(MAGME)]), a photoacid generator ((2,4-dihydroxyphenyl)dimethylsulfonium triflate), and a cross-linker (1,4-butanediol). The development of this resist is conducted in pure water, delivering a competitive 1 μm resolution at a 100 mJ/cm$^2$ sensitivity. Unfortunately, this approach requires a post-exposure backing procedure at 100-125 ℃ to activate the cross-linking process of the exposed areas. This step significantly limits the method compatibility with delicate materials, however keeping a high level of eco-friendliness.

The latter effort resulted in positive and negative tone water-processable resists based on the styrene copolymers,[116] which still required a post-exposure backing for the pattern development. PVA-based resist[117] reported by Wilson's group with (2,4-dihydroxyphenyl)dimethylsulfonium triflate as a photoacid generator is another successful example of the same strategy. Subsequent development included a photoresist based[118] on poly(2-isopropenyl-2-oxazoline) or poly(2-isopropenyl-2-oxazoline-co-styrene) allowed to improve the patterning resolution. The latest negative tone resist[119] with reduced environmental impact is based on poly(1,2:5,6-di-O-isopropylidene-3-O-methacryloyl-α-d-glucofuranose) demonstrated the resolution of about 0.2 μm at 30 mJ/cm$^2$.

Several groups in Korea took up the baton. The first one to mention was a resist based on copolymerization of 4-styrenesulfonic acid sodium salts (SSS) with N-phenylmethacrylamide (copolymer A) or p-hydroxy-N-phenylmethacrylamide (copolymer B) reported by Chae et al.[95] This negative tone resist was utilized with a 254 nm wavelength source and delivered ~1 μm resolution at 1100 mJ/cm$^2$. Further, a water-ethanol mixture soluble resist synthesized by copolymerization of glyceryl methacrylate and methacrolein proposed by Kim et al.[120] was a significant step towards water-developable resist operating with a 193 nm radiation source. As well as the resists developed by Wilson's group, those also required a post-exposure backing procedure. Recent paper by Nothdurft et al.[96] proposed azosulphonate-doped PVA as a negative tone water-developable photoresist, which allowed to eliminate the post-exposure backing procedure. Finally, in 2020 a dextrin-based negative tone photoresist was applied by Toru Amano et al.[121]

Speaking of positive tone resists, we should mention a successful realization of a water-processable resist in 1999 by Guo et al.[122] This research group utilized copolymer Langmuir-Blodgett films of N-(2,2-dimethylpropyl)methacrylamide (DMA) and N-phenylmethacrylamide (PhMA), which delivered 0.75 µm resolution with a Hg lamp exposure. Wilson's group has also reported several positive tone resists. The first one was based on thermal decarboxylation of a half ester of malonic acid,[123] a water-soluble polymer, which however required tetramethylammonium hydroxide (TMAH) as a developer. The second approach employed styrenic polymers,[124] which solubility switches after light irradiation and baking process. Moving on, we should pay attention to a successful example of regular OL resists, such as SU8, with "protective" layers that ensured a high degree of compatibility.[125] In this work the lift-off process (bottom-up approach) was successfully realized(see **Figure 3**) using dextran and polyacrylic acid patterning through the resist. These additional films allowed to keep the substrate intact. In contrast, earlier reports mainly discussed only top-down fabrication.

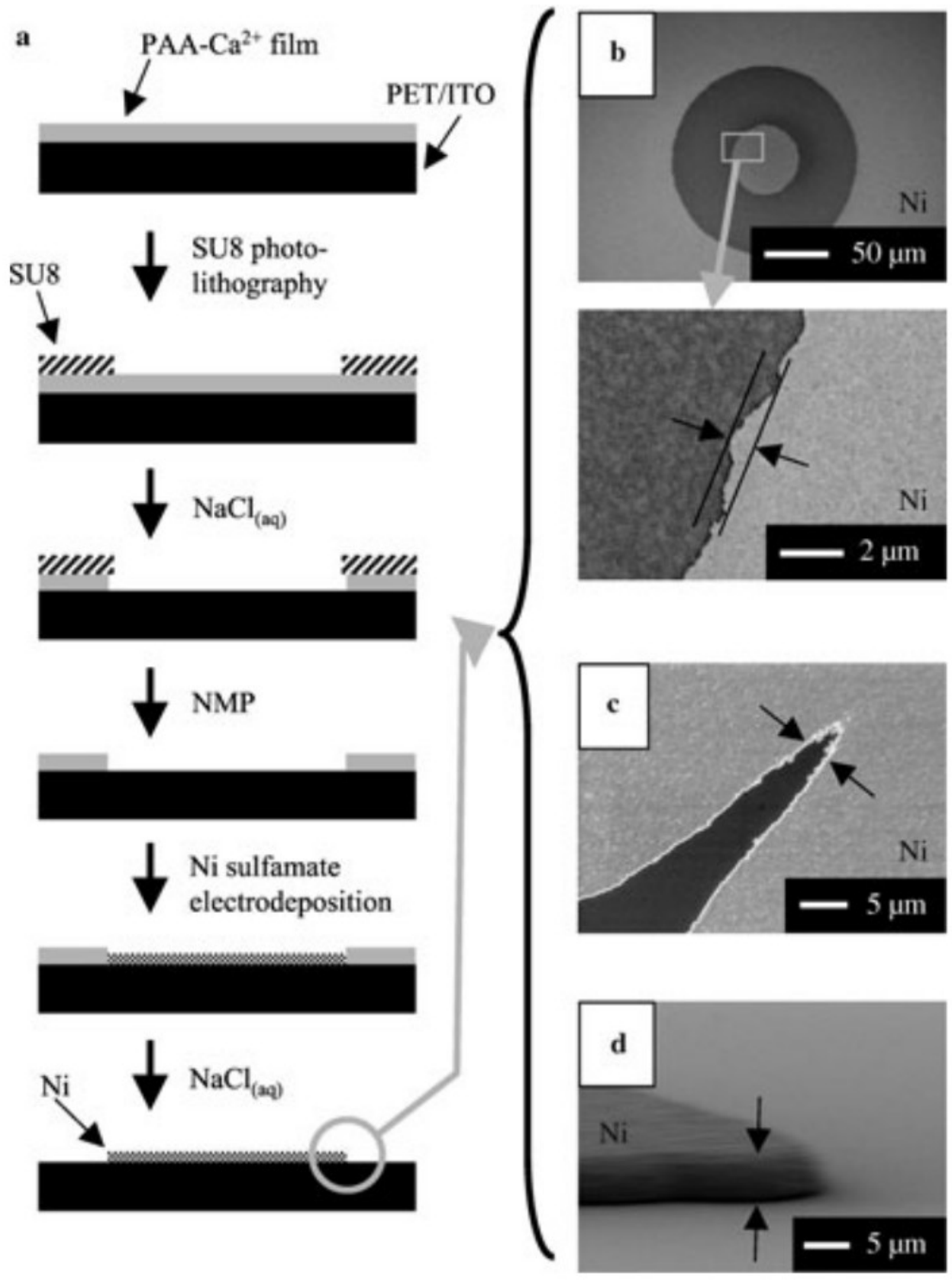

**Figure 3. Electrodeposition of nickel**: a) Schematic illustration of the microfabrication of nickel features by electrodeposition through a film of patterned PAA–$Ca^{2+}$. The substrates of ITO-coated poly(ethylene terephtalate) (PET) conducted the electric current required for the deposition of metal into the openings of the patterned film of PAA–$Ca^{2+}$. b–d) SEM images of electrodeposited nickel on ITO-coated PET. The edge resolution (b) and the lateral resolution (c) of the nickel features was determined by the transparency mask used

for the photolithography. The regions of nickel appear bright in the SEM images. d) The PAA–$Ca^{2+}$ film was stable in the commercial solution of nickel sulfamate for extended periods of time (> 3 h at 40 ℃) during the electroformation of thick nickel features. Reproduced with permission.[125] Copyright 2005, Wiley-VCH.

Among biocompatible techniques, we should point out poly(t-butyl acrylate)-based resist[126,127] with sulfonium salt photoacid generator. These are important works, where the backing temperature and concentration of TMAH developer was lowered to meet the requirements of biological samples. However, the resolution obtained with this resist was relatively poor, just achieving about 5 μm. Kakabakos' group further reported several resists based on (meth)acrylate copolymer[128–130] processed at mild baking conditions: T<50℃ and low developer concentrations.

Recently, Waldbaur et al.[131] proposed an essential technique of biocompatible patterning, which did not require any resist film and employed immobilization of fluorescently labeled molecules and photobleaching. Although this technique focused on protein molecules, baking steps, harsh organic solvents, and bases were eliminated.

It is worth mentioning that a few commercially available resists have been also applied for the patterning of biomaterials. For instance, the hydrophilicity of the baked SU8 resist film changes due to the light exposure, defining a cell adhesion pattern.[132,133]

Generally, the family of optical resists targeting environmental issues and enhanced compatibility is extensive, and many solutions have never been likely published in the form of scientific reports. Table 1 summarizes and systematizes the literature describing different resists, conditions, processing steps, and reagents, including those mentioned above.

### 3.2. Note on compatibility

To differentiate various patterning features in terms of compatibility with bio-organic samples and their effect on the environment, we introduce three-level color grading. We use green to mark a particular processing step or reagent that does not influence a wide range of bio-organic materials or if its influence is reversible. If a technique is mild or fractionally limits the utilization of a resist, we use yellow color. Finally, if the treatment causes massive irreversible changes in the sample material, we assume red.

The first issue is the promotion of covalent cross-linking between sample components, photoresist, and developer material via radical reactions. Exposure of unsaturated carbon-carbon bonds in lipids, proteins, or sensitive co-factors like retinal to ionizing radiation, especially in the presence of oxygen, often leads to irreversible co-polymerization of bio-organic components.[134] However, two primary factors can reduce the negative impact of radical reactions on sensitive bio-organic samples: replacing an oxygen-containing atmosphere with inert during the irradiation and decreasing intermediate radicals' activity during the polymerization of the resist. The absence of molecular oxygen blocks the synthesis of reactive organic peroxides. In its turn, using those chemical compounds in which the spin density is distributed over a larger space can reduce the risk of cross-linking with the sample material. For this reason, polystyrene derivatives appear to be preferable to other substituted polyethylene derivatives such as acrylic acid and polyvinyl alcohol. To avoid radical processes, strong oxidizers should also be excluded.

The heat treatment of samples is the second concern. bio-organic materials can be unstable even at slightly elevated temperatures. A refolding of tertiary or quaternary structures might irreversibly happen in some proteins above 40 ℃. Although thermophilic organisms can exist at 110 ℃ and under high pressure, these are exceptional cases.

It is equally important to mention high ionic strength and significant deviation in pH from neutral values capable of damaging the high-level structural organization of bio-organic systems. On the other hand, dialysis or pH neutralization can make this effect reversible, mainly if the hydrolysis of esters and amides inside the sample material is limited.

We should also pay attention to the physiology of ions used for patterning. A small number of metal cations is basic for most living systems in a fairly wide range of concentrations. These include $Na^{+}$, $K^{+}$, $Ca^{2+}$, $Mg^{2+}$, and $Fe^{2+/3+}$. We always considered their salts with physiological anions (chloride, phosphate, acetate, etc.) in the green zone even though they can cause denaturation of proteins. Essential trace metal cations like $Cu^{2+}$, $Zn^{2+}$, $Mn^{2+}$, $Co^{2+}$, and actively used to bind polyhistidine-tag in proteins $Ni^{2+}$ cation[135] were referred to as the yellow grade.

Finally, we need to mention the drastic influence of utilized solvents. We considered water and dimethyl sulfoxide as non-toxic solvents, i.e., the green grade liquids. Physiological compounds (including those acids arising in the Krebs cycle and glycolysis and their esters or glycerol), all even saturated alcohols and simple ketones, appear in the yellow zone. Ethers (tetrahydrofuran, diethyl ether) turned out to be also in the yellow zone, since they can be efficiently utilized by biological means but can have a significant irreversible effect on the structure of some bio-organic materials. Aldehydes and formic acid are considered to be uniquely toxic. Aromatic solvents (benzene, toluene), dichloromethane, chloroform, carbon tetrachloride, and others fall automatically into the red zone due to their toxicity to the operator and environment. Note that the irreversible decarboxylation of organic materials can be activated thermally and via catalysis by solvent.[136]

**Table 1.** Optical resist and techniques towards enhanced compatibility and environmental friendliness.

Sol - Solvent used to dissolve the resist body material; Pre-exp baking - baking procedure performed after the resist is applied to the surface of target material, but before the exposure; $D_0$ - sensitivity, i.e. minimal radiation dose required for correct resist operation; Post-exp baking - baking treatment required after the exposure process was completed; Dev - developer solution used to define the pattern after exposure procedure; Rem - remover, a solvent or physical process that is used to remove non-irradiated resist film; Res - resolution, either in terms of critical dimension structure or of an individual formed element; BU - bottom-up routine compatibility, usually a lift-off technique is assumed; TD - top-down routine compatibility, usually an etching approach is shown. ”No” stands for the absence of the step. “NI” stands for “not identified”. Colors are used to specify compatibility: red - limited, many bio-organic samples may not withstand this particular step or feature; yellow - mild detrimental processing; green - highly compatible.

| Resist | Tone | Sol | Pre-exp Baking, ℃ (time, min) | $D_0$, mJ/cm$^2$ ($\lambda$, nm) | Post-exp Baking, ℃ (time, min) | Dev (time, s) | Post-dev treatment | Rem | Res, nm | TD | BU | Comment |
|---|---|---|---|---|---|---|---|---|---|---|---|---|
| Poly(MAGME) + PAG + 1,4-butanediol[115] | - | $H_2O$ | 75-90 (1-2) | 100 (248) | 110 (1) | $H_2O$ (30) | No | NI | 500-1000 | + | - | Silicon etching, copper etching. High temperatures cause destruction of the superstructure of bio-organic molecules. Radical is screened by a large substitute. |
| Styrene copolymer/poly acrylic acid[116] | -/+ | $H_2O$ | 130 (1-2) | ~200 (254) | 90 (1) | $H_2O$ (30-60) | No | NI | 1000-2000 | + | - | Silicon etching, copper etching. High temperatures cause destruction of the secondary structure of bio-organic molecules. Radical is not screened by a large substitute in acrylic acid monomers. |
| Poly(vinylalcohol) with 2,4-dihydroxyphenyl) dimethylsulfonium triflate as PAG and hexamethoxymethyl-melamine as crosslinking agent[117] | - | $H_2O$ | 120 (3) | 200 (254) | 100-130(2) | $H_2O$ (2x 10) | No | NI | 1000-1500 | + | - | Silicon etching, copper etching, phosphor raster screens, PCB, CCD displays. High temperatures cause destruction of the superstructure of bio-organic molecules. Radical is not screened by a large substitute. |
| Poly(2-isopropenyl-2-oxazoline) or poly(2-isopropenyl-2-oxazoline-co-styrene) and PAG (4-Methoxyphenyl)-dimethylsulfonium-triflate[118] | - | $H_2O$ | 120 (3) | 65 (254) | 120 (2) | $H_2O$(30) | No | NI | 1000-2000 | + | - | High temperatures cause destruction of the superstructure of bio-organic molecules |
| Poly(1,2:5,6-di-O-isopropylidene-3-O-methacryloyl-R-D-glucofuranose) and PAG (4-Methoxyphenyl) dimethylsulfonium Trifluoromethanesulfonate[119] | - | $H_2O$ | 120 (3) | 30 (248) | 110 (2) | $H_2O$ (30) | No | NI | 200 | + | - | High temperatures cause destruction of the superstructure of bio-organic molecules |
| (SSS) copolymer with p-hydroxy-N-phenylmethacrylamide)[95] | - | $H_2O$ | 120 (20) | 1100-26600 (254) | No | $H_2O$ (20) + acetone wash | 120 (2 min) | NI | 1000 | + | - | High temperatures cause destruction of the superstructure of bio-organic molecules. An active radical initiator is used for the photoresist copolymerization (AIBN). |
| Glyceryl methacrylate and methacrolein[120] | - | $H_2O$: EtOH =8:2 | prebake unknown conditions | 45 (195) | 130 (1.5) | $H_2O$ | No | water-ethanol | 700 | + | - | High CF4 dry etch resistance.<br>High temperatures cause destruction of the super structure of bio-organic molecules. |
| AZO doped PVA[96] | - | $H_2O$ | 100 (5-60) | 740*60 (Hg) | No | $H_2O$ | No | NI | 10000-20000 | + | - | High temperatures cause destruction of the secondary structure of bio-organic molecules.. The tendency of azo compounds towards radical reactions softened by aqueous environment. |
| Dextrin[121] | - | $H_2O$ | 120 (1) | 600 (Hg) | No | $H_2O$ (90) | No | NI | 1000 | + | - | High temperatures cause destruction of the superstructure of bio-organic molecules. |
| DMA and PhMA[122] | + | *LB film | No | Hg | No | $H_2O$, 35 ℃ (30) | No | NI | 750 | + | - | Radical is screened by a large substitute. |
| Thermal decarboxylation of a half ester of malonic acid[123] | + | $H_2O$ | 165 (4) | 20 (248) | 140 (1) | TMAH (10 s) | No | NI | 1000 | + | - | High temperatures cause destruction of the superstructure of bio-organic molecules. |
| Pendant ammonium salts of half-esters of malonic acids and acid-labile alkyl esters[124] | + | $H_2O$ | 165 (5) | 30 (248) | 140 (5) | TMAH (30) | No | NI | 600 | + | - | High temperatures cause destruction of the superstructure of bio-organic molecules. |
| PAA/Dextran/PMAA + S1811 photoresist[125] | + | $H_2O$ * protective layer | 100 (2) Typical for S1811 | 150 (254) | No | S1811 developer + $H_2O$/ 1M NaCl + 0.05% Tween 20 | No | $H_2O$ | 1000-3000 | + | + | Technique of compatibility enhancement by means of addition of protective layer. Ensures bottom-up and top-down approaches. Good solution for bio-organic systems. Needs a slight decrease in temperature. High temperatures cause destruction of the supersecondary structure of bio-organic molecules. |
| Poly(t-butyl acrylate) with sulfonium salt photoacid generator[126] | + | ethyl lactate/toluene | 60 (5) | 11 (254) | 60 (5) | $2.7\times10^{-3}$ M TMAH | No | $2.7\times10^{-3}$ M TMAH, and additional UV curing | 5000-10000 | + | + | Bottom-up and top-down patterning of biomolecules. Toluene is toxic and can disrupt hydrophobic regions of bio-organic materials. Radical is not screened by a large substitute. |
| Methyl acrylate-based[127–129] | +/- | toluene | 50 (2) | 36-106 (254) | 50 (2) | $13\times10^{-3}$ M TMAH | No | $13\times10^{-3}$ M TMAH | 130 | + | + | Improved resolution in comparison for poly(t-butyl acrylate) based resist. Toluene is toxic and can disrupt hydrophobic regions of bio-organic materials. Radical is not screened by a large substitute. |
| Maskless projection lithography for protein patterning[131] | +* | No mask | No | White light within 10 s | No | No | No | No | 2500 | + | - | * - specific for fluorescently labeled molecules |
| Indomethacine[137] | - | VD* | No** | 200 (254) | No | 0.002N TMAH | No | NI | 1000 | - | - | * Also a solvent based route has been shown. VD requires low temperatures 265 K<br>** Solvent based route requires 100 (1) baking |

### 3.3. Electron beam lithography

Electron and ion beam[138] lithography (EBL) differs from OL by the radiation source, while the principle of a pattern creation is the same. A resist film (usually polymeric one) acts as an intermediate layer,[139] but most OL resists are incompatible with electron and ion radiation. For all materials reviewed below, we indicate the essential technological parameters: sensitivity, i.e., the minimal dose required for complete development, which strongly depends on the accelerating voltage indicated in parentheses; resolution; and various additional pre- and post-exposure routines.

For EBL, many bio-inspired natural resist materials have been already reported, while the diversity of OL resists is mainly limited to synthetic ones. That is essential, as bio-derived solutions ensure excellent compatibility with bio-organic materials and are eco-friendly in terms of material synthesis and waste production. Another significant advantage of EBL is that this method delivers unprecedented high resolution when compared to OL (in regular laboratory facilities).

Of course, the EBL method is not without drawbacks. The most important one is the interaction of bio-organic matters with an electron radiation, which is more destructive when compared UV-visible range waves. The second disadvantage is the requirement of high vacuum conditions for the best operation. In contrast, OL is usually performed under ambient conditions. The third EBL drawback is a long exposure time, which can be several orders of magnitude longer when compared to OL. However, often EBL is the only option for most laboratory nanoscale fabrication.

Common e-beam resists (e.g., PMMA) require pre-exposure baking at 150 ℃ and inevitable contact of the substrate with strong organic solvents: anisole (resist solvent for casting), methyl isobutyl ketone/isopropanol (development), and acetone (removal). Such e-beam resists demonstrate high resolution around 5-10 nm at the sensitivity of a few hundreds $\mu C/cm^2$ (at 30-50 kV).

The use of bio-organic materials, primarily proteins, polypeptides, polysaccharides (including amino saccharides such as chitosan) as a resist is associated with various advantages. Many of these materials, when scaled up, turn out to be cheaper than materials obtained by chemical synthesis. These materials and their processing products are biodegradable and biocompatible. Water is the primary medium for the resist dissolution before or after radiation exposure. It is worth mentioning that the solubility of the bio-organic materials in water depends on relatively weak factors. Here, we should mention a slight change in the pH of the medium and the effect of physiologically compatible metallic ions. Physical treatments include a small dose exposure of the ionizing radiation and slight overheating within 30-40 °C above room temperature. Such mild processing conditions are usually more important than a slight loss of spatial resolution when compared to contemporary resists.

Let us first review several cutting-edge techniques reported in the past decade, focusing on bio-inspired solutions, as they meet the requirements motivated our survey: environmental friendliness and enhanced compatibility. Next, we mention several unconventional techniques that utilize the properties of conductive organic polymers and water-soluble inorganic materials. Finally, we give a remark on an ice lithography approach.

### *3.3.1. Protein-based resists*

The first subgroup we review here is protein-based resist materials. Remarkably, barely all reported solutions exhibited a positive tone behavior. Most of them were even capable of tone-switching by means of various additional chemical and physical treatments, keeping the sensitivity dose within the same order of magnitude. We start with silk fibroin - a natural or recombinant protein.[140–142] Although preparation of aqueous fibroin solutions assumes dialysis procedure, all other steps meet standards of regular EBL. Silk-based resist supports switching between negative and positive tones by inducing conformational changes in the protein molecules. Water-soluble alpha spirals form a body of the negative tone resist, while water-insoluble beta-sheets are used as the positive tone. Interacting with beta-sheets, electron beam radiation destroys its crystalline phase and turns the irradiated areas into the water-soluble state.

On the other hand, when an e-beam hits alpha-spirals, it induces cross-linking, inhibiting further dissolution. The transformation from alpha-spiral to beta-sheet is induced by processing with methanol or water steam treatment. Even though the formed pattern is not entirely suitable for a lift-off application (see **Figure 4**), it delivered 50 nm resolution with roughly 2000 μC/cm$^2$ sensitivity (at 125 kV) compatible with etching routines. Modification of fibroin molecules allowed further utilization of negative patterns for a fluorescent microscopy.

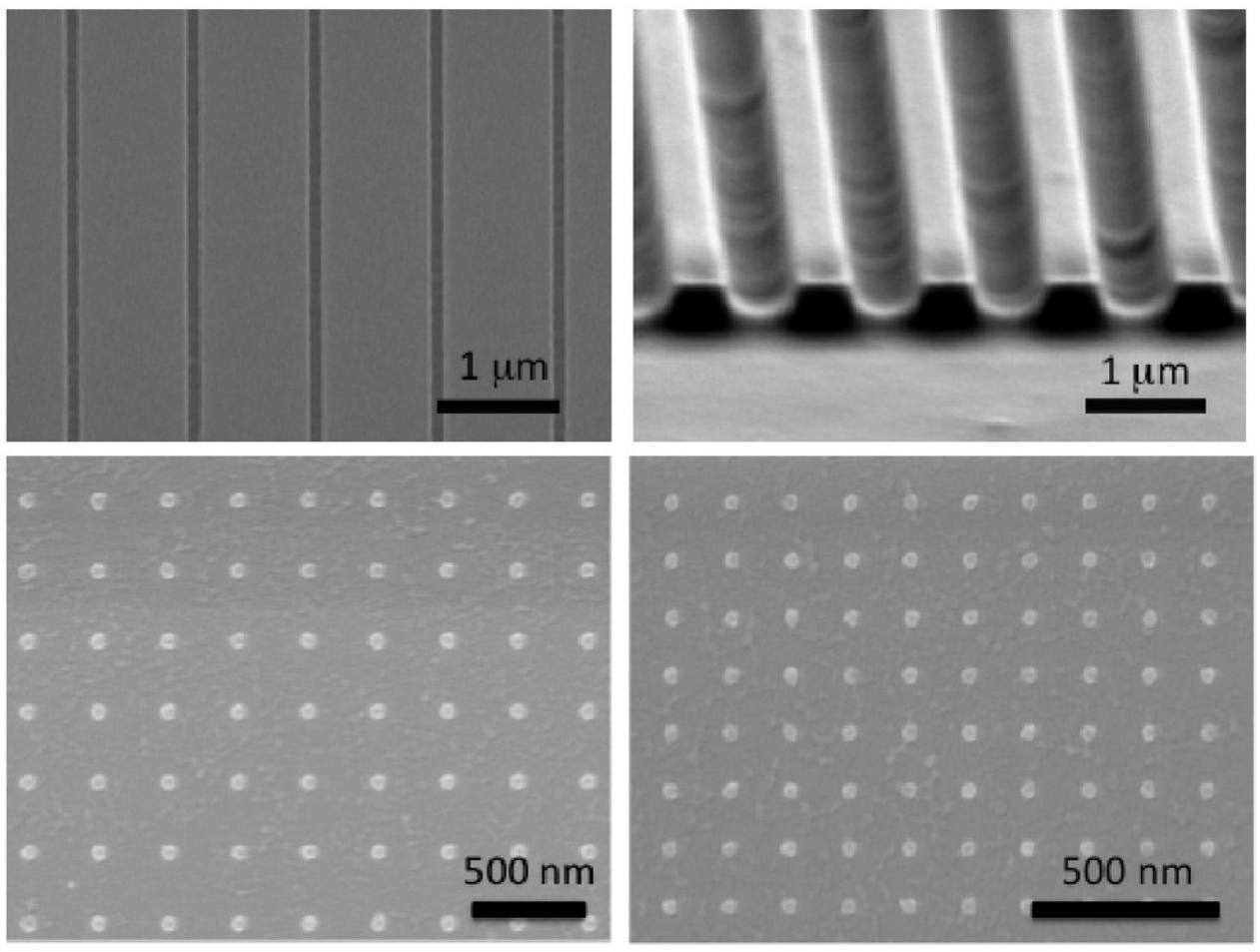

**Figure 4. Test structures fabricated by fibroin lithography.** SEM image of the 1D line pattern with the 100-nm width (top) and tilted view with the 500-nm line width pattern showing 1:1 aspect ratio. 2D test structures (bottom) showing 50 and 30 nm pillars with 150 and 250 nm lattice constant at 1:1 aspect ratio. Reproduced with permission.[140] Copyright 2014, Nature Publishing Group

Another successful realization of the protein-based EBL resist[143] is a chicken egg protein originally derived in liquid form and available in barely any local supermarket. As well as in the case of fibroin, a dual-tone processing is possible when employing an additional step - UV curing. An etching process with a template on a substrate surface was demonstrated with sub-100 nm resolution at 3000 μC/cm$^2$ (30 kV).

A bright and deeply investigated example is based on lysozyme phase transition, when mixing with tris(2-carboxyethyl)phosphine (TCEP) buffer at neutral pH.[144] The film formed due to this phase transition can be transferred to an arbitrary substrate by the Langmuir-Blodgett method. The authors demonstrated numerous applications with a high resolution of 50 nm and sensitivity of 2500 μC/cm$^2$ (30 kV). An essential feature of this method is that no pre-exposure treatment of the resist film is required to process the patterning.

Recently a high-resolution 3D e-beam writing has been demonstrated with a fibroin relative, recombinant spidroin protein, utilized as a resist.[145] However, the authors mainly focused on the formulation of non-planar structures, and they did not examine traditional routes. In all above-mentioned techniques, the developer was just pure water.

### *3.3.2. Natural polysaccharide-based resists*

The following group of naturally derived materials serving as a resist is natural polysaccharides. The first particular solution to mention is a sugar-based negative tone resist[146] that facilitated water-based development of the resist film with a 100 nm resolution at 18 μC/cm$^2$ (30 kV). Another example, natural linear polysaccharide pullulan,[151] demonstrated a resolution of 100 nm at a sensitivity of 10 μC/cm$^2$. This negative tone resist is compatible with both EUV and e-beam exposure. In contrast to protein resists discussed earlier, sugar-based ones require much less radiation dose in negative tone solutions.

Actual progress in the design of positive tone resists was achieved using chitosan[148,149]. Historically, Voznesenskiy et al.[148] published the first report, demonstrating positive tone e-beam resist with a 100 nm resolution and sensitivity of 150 μC/cm$^2$ (at 25 kV). Further research by Calillau et al.[149] revealed the possibility of chitosan application for top-down fabrication through patterning of 50 nm lines in silica (with approximately the same sensitivity). One of the milestone accomplishments is the decrease of required radiation dose compared to earlier reported naturally derived resist materials.

Modified chitosan molecules, i.e., chitosan derivatives, were next reported to be capable of both bottom-up and top-down approaches. This water-soluble and water-developable resist facilitates 100 nm reproducible resolution when fabricating individual metallic lines by the lift-off lithography.[150] This approach utilizes specific chemical reactions of chitosan molecules - the formation of chelate complexes with transition metal ions, which inhibit the solubility of non-irradiated film regions. As a result, the development process is a competition of two opposite reactions - disentanglement and "fixation" of polymer film, and these reactions are sensitive to the polymer molecular weight. The latter is changed by e-beam radiation. Chitosan-derivatives-based method demonstrates the sensitivity of 130 μC/cm$^2$ (at 50 kV). The authors demonstrated direct sub-micron lift-off patterning of organic semiconductors, namely rubrene and tetracyanoquinodimethane (TCNQ) (see **Figure 5**). Moreover, this method allowed creating electronic devices with bio-organic and inorganic parts, including patterning directly on an individual brain microtubule with Pd contacts.

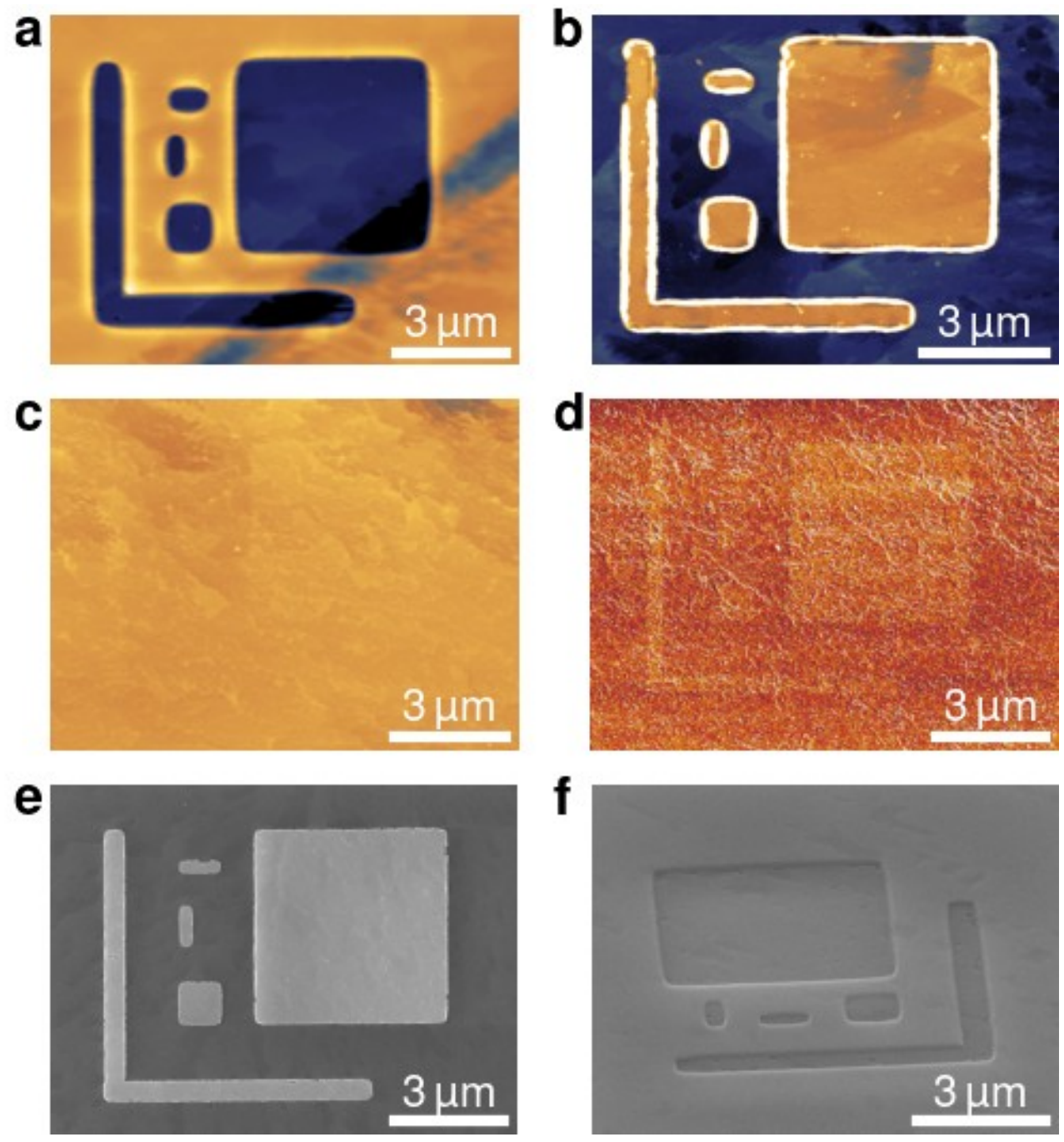

**Figure 5. Lithographic patterning directly on the surface of organic semiconductors by means of chitosan derivatives based resist.** a) AFM image of the developed area in the resist on the surface of a TCNQ crystal. b) AFM image of a similarly processed region of TCNQ crystal after the deposition and lift-off of a Pd film. c,d) AFM topography and phase images of the surface of TCNQ single crystals exposed to the same doses as in (a) and (b) and processed by developer solution. SEM images of e) lifted-off Pd structures and f) oxygen plasma etched pattern fabricated on the surface of a rubrene thin film. For all presented patterns, ≈100 nm resist film and irradiation dose of ≈2000 $\mu Ccm^{-2}$ were used. Reproduced with permission.[150] Copyright 2021, Wiley-VCH.

Another example of polysaccharides, dextrin derivatives[151], demonstrated a high resolution (20 nm) positive tone ethanol-developable resist behavior with the sensitivity of roughly 1800 $\mu C/cm^2$ (125 kV). However, the obtained derivative was not water-soluble, which limited application towards bio-organic materials. Biomass-derived negative tone resist based on alpha-linked disaccharide[152] with an e-beam sensitive acrylic group was utilized for plasma etching patterning with an organic layer underneath. The utilization of the additional film and rather aggressive resist removal procedures do not allow us to claim this solution highly compatible, but it is an essential step to eco-friendly production of microelectronic devices.

Finally, we underline an exciting application of trehalose[153,154] for direct-write protein patterning. The saccharide layer acts as the negative tone resist and a stabilizing agent for an encapsulated protein. Trehalose layer crosslinks under e-beam radiation and forms a negative tone pattern. At the same time, its specific properties help save encapsulated proteins intact. Therefore, final properties of obtained resist structures are specified by the reagent fixed inside – proteins, like e.g. enzymes.

*3.3.3. Non-bio-derived resists*

Here we move on to some synthetic polymers[139] that are also capable of a higher degree of compatibility when compared to commercially available solutions. The conductive water-soluble polymers, like, e.g., polyaniline,[155] can be used to form a negative tone pattern. Therefore, it can be considered as an intermediate solution to create conductive lines substituting traditional lift-off. In 1993, the IBM T.J. Watson Research Center group reported both the resist material synthesis and corresponding patterning technique. Two decades later, Abbas et al. reported a peculiar utilization of polystyrene sulfonate as a negative tone resist with remarkable lift-off capability[156]. We should also mention water-soluble negative tone resists based on fullerenes.[157] The authors showed high-resolution (30 nm) negative tone patterning with an acceptable sensitivity of 800 μC/cm$^2$, eliminating pre-exposure or post-exposure treatments.

As well as for OL, PVA has been often utilized as a resist basis in various blends with metals salts, like LSMO[158] and nanoparticles[159]. PVA easily cross-links under e-beam radiation, facilitating negative tone resist solutions for planar and 3D e-beam lithography. It is worth mentioning that these approaches aim to obtain a composite structure from the resist on the surface of the target material. The authors however did not pursue the utilization of the resist as a mask for subsequent patterning.

In the case of synthetic polymers, there is an interesting report on poly(glycidol) nanogels[160], similar to trehalose-based resist[153,154]. Nanogel acts as a negative-tone resist and a stabilizing agent for encapsulated proteins, particularly for patterning the three-enzymes cascade. **Figure 6** illustrates the operation flow for this approach.

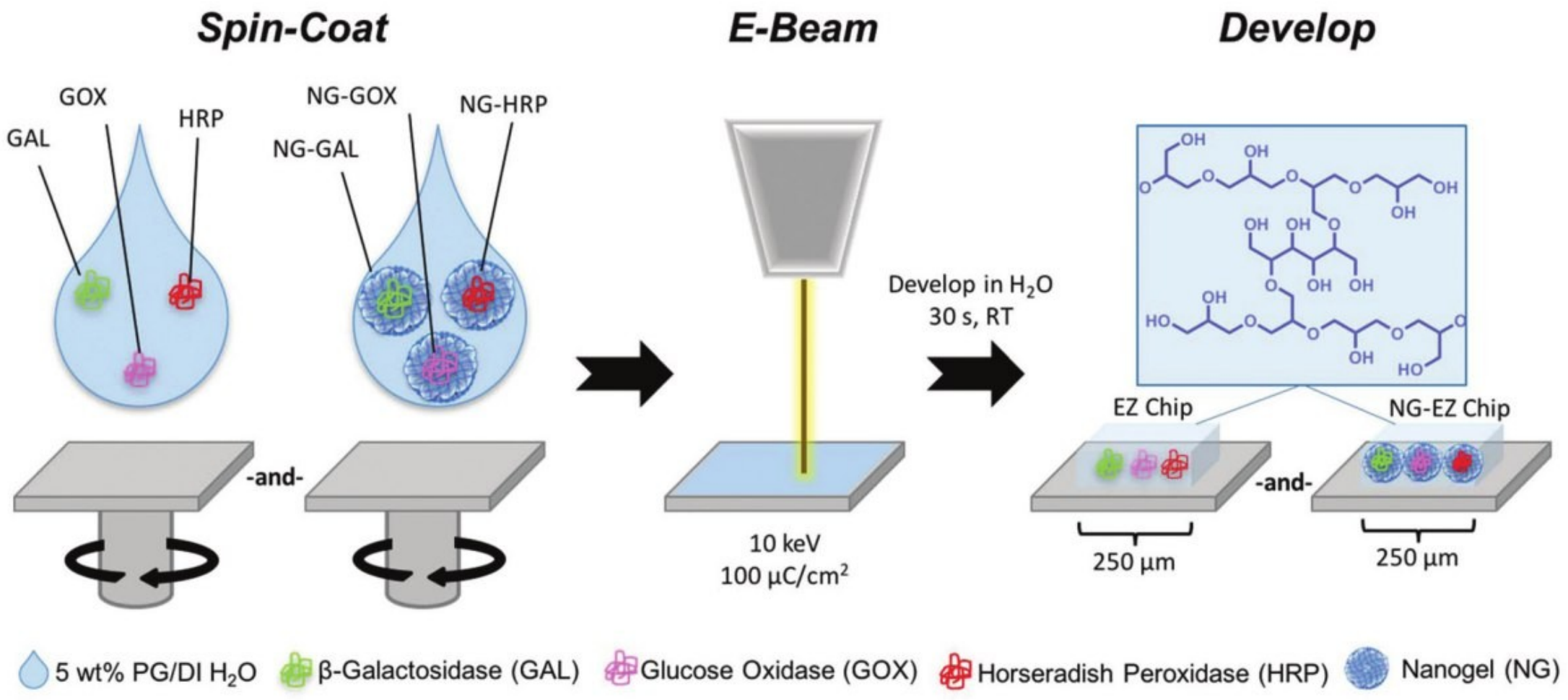

**Figure 6.** Device fabrication from spin-coated nanogels (NG) loaded with enzyme (NG-EZ) or free enzyme (EZ) within an aqueous solution of 5% semi-branched poly(glycidol) on piranha-cleaned silicon chips. Electron beam lithography (EBL) was employed to cross-link poly(glycidol) in user-defined 250 μm square patterns. Following EBL, chips were developed in water to wash away non-crosslinked polymer material to yield 3-D

hydrogel structures of immobilized enzymes. Reproduced with permission.[160] Copyright 2018, Royal Society of Chemistry

### 3.3.4. *Ice lithography*

A distinct group to mention is so-called ice lithography (IL)[161] using water[162,163] for positive tone patterning, and solid alkanes[164,165] and recently reported anisole[166] for a negative one. Unlike methods mentioned earlier, IL skips steps of resist coating and development in terms of utilization of any additional solvents. Instead, the resist material is vapor-deposited from a gas phase under cryogenic temperatures. For positive tone solutions, the resist self-develops by evaporation during e-beam exposure. For the negative one, irradiated areas cross-link and remain after the temperature rise and evaporation of intact resist. These techniques facilitate extremely high resolution up to 5 nm and allow lab-scale demanded lift-off techniques. A particular example of the utilization of ice for the bottom-up creation of Pd contacts is illustrated in **Figure 7**.

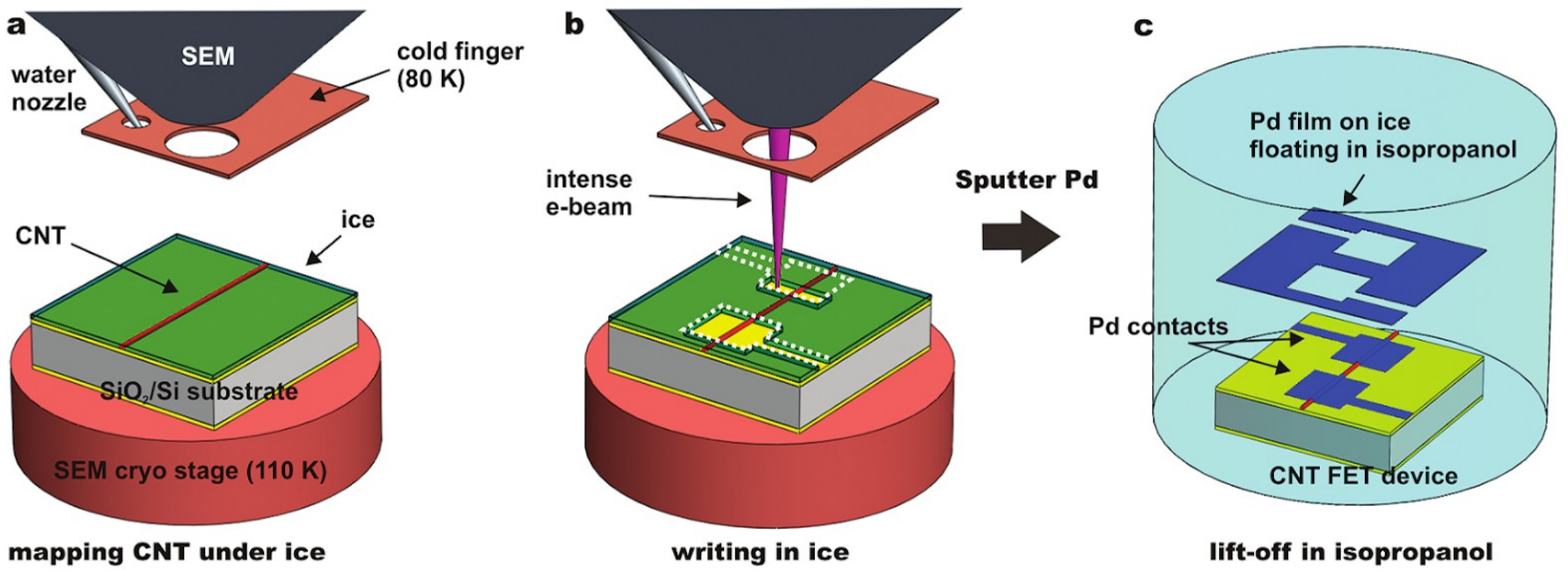

**Figure 7. Ice lithography process.** (a) The sample with preformed Mo microleads and SWCNT on the $SiO_2$-coated Si substrate is loaded into the SEM via the load-lock and cooled down to ~110 K on the SEM cryo-stage. Water vapor is leaked into the SEM through a nozzle just above the sample and condenses as amorphous ice on the cold sample. Typically, 80 nm of ice is deposited in 30 s. The location of a SWCNT under the ice is mapped. (b) An intense e-beam draws patterns for the contacts (white dotted line) and removes ice, forming a mask for metal electrodes contacting the SWCNT. (c) The sample with nanopatterned ice resist is transferred onto the metal deposition chamber, and Pd is sputtered over the entire sample. The sample is removed from the metal deposition chamber and, while still frozen, immediately immersed into 2-propanol held at room temperature, whereupon the Pd film on top of the ice resist drifts away, leaving the preformed Mo leads connected to the SWCNT with Pd interconnections only where the ice had been removed by the e-beam. Reprinted with permission.[162] Copyright 2010 American Chemical Society.

Cryogenic temperatures limit the application of IL with some bio-organic samples. However, in the case of a water-based approach, the authors claim the formed ice to be amorphous, resulting in the reduced damage. Many synthetic materials can withstand low temperatures, and overall, IL assures a high level of eco-friendliness. Even though EBL systems capable of such patterning types are not widespread at the moment, they represent a promising solution that may grow to a commercial level one day.

All EBL techniques mentioned above, particular features, processing steps, and utilized reagents are summarized in **Table 2**.

**Table 2.** EBL resists and patterning techniques towards enhanced compatibility and environmental friendliness.

Sol - Solvent used to dissolve the resist body material; Pre-exp baking - baking procedure performed after the resist is applied to the surface of target material, but before the exposure; $D_0$ - sensitivity, i.e. minimal radiation dose required for correct resist operation; Post-exp baking - baking treatment required after the exposure process was completed; Dev - developer solution used to define the pattern after exposure procedure; Rem - remover, a solvent or physical process that is used to remove non-irradiated resist film; Res - resolution, either in terms of critical dimension structure or of an individual formed element; BU - bottom-up routine compatibility, usually a lift-off technique is assumed; TD - top-down routine compatibility, usually an etching approach is shown. ”No” stands for the absence of the step. “NI” stands for “not identified”. Colors are used to specify compatibility: red - limited, many bio-organic samples may not withstand this particular step or feature; yellow - mild detrimental processing; green - highly compatible.

| Resist | | Tone | Sol | Pre-exp Baking, ℃ (time, min) | $D_0$, μC/cm$^2$ (AV, kV) | Post-exp Baking, ℃ (time, min) | Dev (time, s) | Post-dev treatment | Rem | Res, nm | TD | BU | Comment |
|---|---|---|---|---|---|---|---|---|---|---|---|---|---|
| Fibroin - natural silk peptide[140–142] | Peptides and proteins | +/-* | $H_2O$** | 90 (1) | 2250/25000 (100) | No | $H_2O$ (60) | Plasma | NaCl, LiBr | 30 | + | +.- *** | * - methanol/water vapor treatment switches tone. ** - dialysis is required. *** - poor lift-off quality |
| Spidroin[145] | | + | $H_2O$ | 110(30) | 20-20000 (0.5-10) | No | $H_2O$(30-300) | No | NI | 15 | + | - | Standard routines not shown, more applicable for micro- and nano-fluidics |
| Chicken egg protein[143] | | +/-* | No | UV precuring 1200 mJ/cm$^2$ for positive | 3000/1500 (30) | No | $H_2O$ (30) | Plasma | 0.25% trypsin at 37deg/ 8M urea 12-15 hours | 15 | + | - | * - UV switchable tone Also compatible with 254 nm UV lithography |
| lysozyme[144] | | + | * | No | 2500(30) | No | $H_2O$ (60) | No | 3 M of guanidine | 50-100 | + | - | * - Langmuir-Blodgett film |
| | | | | | | | | | hydrochloride aqueous solution/50% acetic acid aqueous solution | | | | deposition |
| Chitosan[148,149] | Carbohydrates and derivatives | + | 0.7-2.5 % Acetic Acid | Treatment with 3 % ammonia solution for 10 min/100 (1) | 150-160 (25) | No | H2O (30) | Ar ion etching | Oxygen Plasma | 50 | + | - | |
| Chitosan Derivatives[150] | | + | $H_2O$ | No | 130-2000 (50)* | No | TMS** aqueous solution (300-600) | No | Low concentrated organic acids (0.01-0.1%) and $H_2O$ | 50-100 | + | + | * - Sensitivity depends on the developer concentration. Top-down and bottom-up routines. Complete development.<br>** - Transition metal salt (TMS) |
| Polystyrenyl ether trehalose[153] | | - | $H_2O$ | No | 5-80 (30)* | No | 1 mM Tween 20 in D-PBS | No | No | 80-100 | - | - | * - Sensitivity depends on the encapsulated protein. |
| Polyaniline[155] | Polyaromatic ionogenic | - | $H_2O$ | 80 (3) | 200 (50) | 80 (5) | $H_2O$ (dip) | IPA rinse | No | sub-1000 | - | - | Conducting lines made of crosslinked resist can be created |
| Poly(sodium 4-styrenesulfonate)[156] | | - | $H_2O$ | 90 (5) | > 2000 (20) | No | $H_2O$ (10) | No | $H_2O$ | 60 | + | + | Patterning of organic semiconductors and conductive polymers |
| fullerol[157] | Other | - | $H_2O$ | No | 850-15000 (20)* | No | $H_2O$ (10) | No | Oxygen plasma | 30 | + | - | *Higher sensitivity cam be obtained by chemical amplification and requires postexposure bake |
| Water ice[162,163] | Ice lithography | + | VD | Cryo-T | $10^6$ | No | No | No | Isopropanol | <10 | + | + | Unconventional SEM. Utilization of cryo temperatures can be compatible and soft towards biorganics if formation of hexagonal ice is avoided. |
| Organic ice resist[164–166] | | - | VD | Cryo-T | 3000-12000 | No | No | No | No | <10 | + | - | Unconventional SEM. Utilization of cryo temperatures can be compatible and soft towards biorganics if formation of hexagonal ice is avoided. |

We will briefly further review some other unconventional lithographic methods. They often outperform EBL and OL due to, e.g., resist film elimination or much softer physical interaction, but not widely spread, often not scalable, and with patterning field size limitations or non-precise localization.

### 3.4. Multiphoton lithography

First to mention is multiphoton lithography,[167,168] where during direct laser writing on the surface of polymer resist (the negative or positive one), two-photon absorption (or even higher-order processes) result in modification of the resist material property (such as solubility) . We intentionally set this section apart from regular OL, as it often assumes home-built lithography set-ups and custom resist materials. Several groups have shown successful patterning of bio-organic objects with the help of multiphoton lithography fabricating various 3D scaffolds.[142,167] Perevedentsev et al.[169] showed a rather unconventional path towards patterning molecular electronic devices utilizing laser radiation. The authors achieved a photolithographic resolution of less than 5 μm. They achieved a local diffusion of functional small-molecular compounds through a solution-processed 'molecular gate' interlayer.

### 3.5. Nanoimprint lithgoraphy

Nanoimprint lithography (NL)[170] employs special "stamps" created by other lithographic techniques, usually EBL or OL. NL stamps are further applied to the target material surface to transfer the defined templates. Often NL requires intermediate films - resist - which can be polymeric or inorganic. This method often uses additional physical treatments like heating or UV curing to activate additives or fix the resist layer. Cellulose derivatives[171] are an excellent example of water processable biocomaptible NL resists. Direct NL patterning of protein molecules is also actively explored.[172,173] An unconventional method, the so-called "chemical lift-off" approach, allows precise removal of a polymeric film from the material surface, opening it for contacting.[174] NL techniques often reach the high-resolution and excellent degree of eco-friendliness. However, the positioning of the stamp relative to the individual nano-object is quite limited, and most lithography stations are custom.

### 3.6. Dip-pen lithography (Atomic Force Microscopy lithography)

The utilization of AFM probes for local modification of substrates formed a particular group of methods - Atomic Force Microscopy Lithography (AFML). Such techniques are an established key solution for the fabrication of the state-of-the-art samples in 2D van der Waals heterostructures[175] and electrochemical modification of semiconductors for quantum point contacts formation.[176-178] AFML possesses excellent resolution properties <40 nm, great localization precision, and allows high-quality, non-destructive simultaneous visualization. Electrochemical AFML, for example, finds its use for delicate material patterning.[179] Its operation principle implies the interaction of the AFM tip with a chromium layer deposited on top of the polymer films. This approach delivers 9 nm features in the top metal layer and 40 nm lines in the polymer film. Even less harmful method, i.e., thermochemical scanning probe lithography,[180] employs local mild heating by an AFM probe. It resulted in sub-10 nm patterning with high throughput for the fabrication of enzyme nano-patterns. It is worth mentioning that even simple tip oscillations can be utilized for patterning as has been already demonstrated by the tip-driven ultrahigh-resolution patterning of biological molecules in the liquid.[181]

## 4. Conclusion

Despite the tremendous efforts made to create resists for electron-beam and optical lithography compatible with bio-organic materials, there is always room for improvement. Material Science, Biophysics, and Medicine formulate novel challenges that require the development of tools, methods, and employed chemicals.

The most mature industrial technology, optical lithography, still lacks the resist that would not be harmful to various objects, including biological systems (bacterial nanowires, DNA, artificial peptide, etc.), modern organic semiconductors (e.g., rubrene, TCNQ), and even some inorganic delicate materials. The same is true regarding lab-scale e-beam lithography patterning. Further, resist material design would help shed light on the origin of different fundamental phenomena occurring at the nanoscale in delicate systems. Meanwhile, some particular materials may already be patterned thanks to the solutions reviewed in our survey. Soft patterning techniques employing features of biological materials, such as peptides and natural polysaccharides, deliver an unprecedented level of resolution and compatibility.

Radiation-based lithography might help overcome limitations of contemporary implantable medicine, therapeutic and surgical actions. Implementation of submicron patterning on the surgical operation site requests soft, non-toxic materials and techniques to replace commercial resists. It is impossible to overestimate the potential of this method regarding neurosurgery, TENGs, and various stimulators.

We should also mention environmental aspects, as the microelectronics industry produces numerous e-wastes. The creation of eco-friendly resists and techniques is a key to lowering toxic emission and making the production line safer for the operators working with resist solutions and developers. The utilization of natural materials seems to be the most straightforward solution from the environmental point of view.

**Acknowledgements**

A. K. G. and K. A. M. acknowledge funding from Russian Science Foundation under grant 19-73-10154. A. G. N. acknowledges Russian Foundation of Basic Research project no. 20-03-00804. We acknowledge the Ministry of Science and Higher Education of the Russian Federation (project no. FZSR-2020-0007 in the framework of the state assignment no. 075-03-2020-097/1).

Received: ((will be filled in by the editorial staff))
Revised: ((will be filled in by the editorial staff))
Published online: ((will be filled in by the editorial staff))

**Dr. Artem K. Grebenko** received his PhD degree from Moscow Institute of Physics and Technologies, Russia in 2021. He is currently a postdoctoral researcher at Skolkovo Institute of Science and Technologies working on patterning unconventional delicate materials and low-dimensional materials physics.

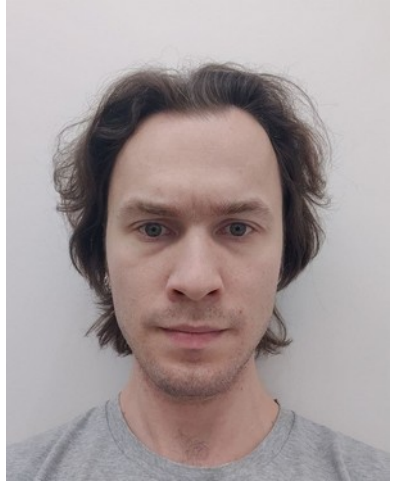

**Dr. Konstantin A. Motovilov** received his PhD degree from the Department of Chemistry of Moscow State University, Russia in 2009. Currently he leads the biophysical group at the terahertz spectroscopy laboratory of Moscow Institute of Physics and Technology. The major area of his interest is the relationship between the mesoscopic organization of water and the processes of electron and proton transfer in bio-organic systems.

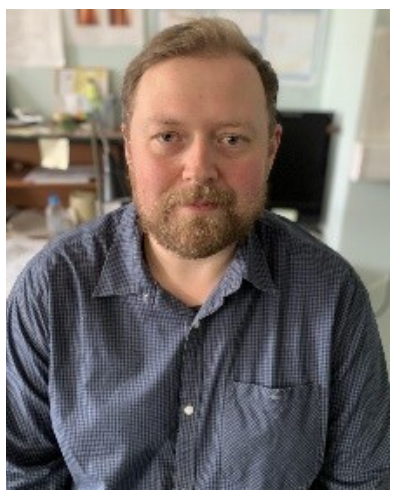

**Anton V. Bubis** received his MSc degree from Moscow Institute of Physics and Technology in 2017. He graduated from Skolkovo Institute of Science and Technology in physics in 2021. He fabricates and studies electronic transport in devices exhibiting topological phenomena at ISSP RAS, Russia.

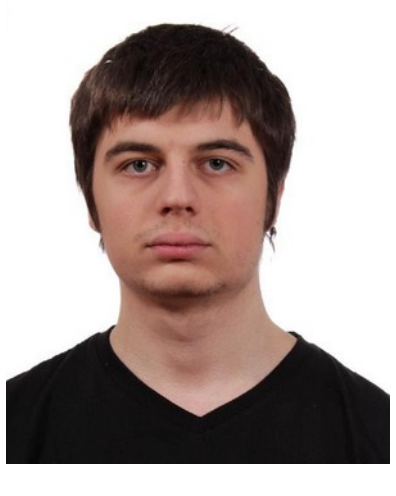

**Dr. Sc. Albert G. Nasibulin** is a Professor at Skolkovo Institute of Science and Technology and an Adjunct Professor at the Department of Chemistry and Materials Science of Aalto University. He got his PhD in Physical Chemistry (1996) at Kemerovo State University (Russia) and Doctor of Science (Habilitation, 2011) at Saint-Petersburg Technical State University (Russia). Prof. Nasibulin specializes in the aerosol synthesis of nanomaterials (nanoparticles, carbon nanotubes and tetrapods), investigations of their growth mechanisms and their applications.

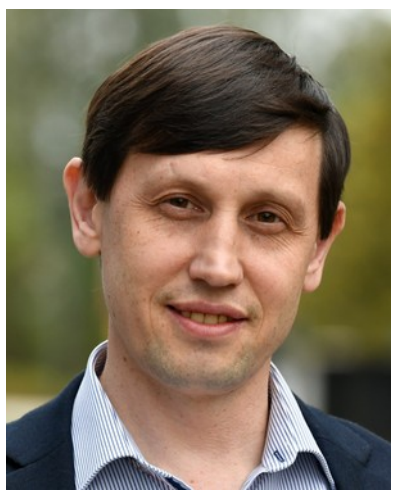

ToC Entry: Micro- and nano-scale patterning is a vital instrument for contemporary science and industry. The emergence of novel delicate materials and the development of biotechnology raised interest in gentle patterning techniques. This review is devoted to the manifold of e-beam and UV resists, direct-write techniques, and non-radiation-based approaches facilitating enhanced compatibility and environmental friendliness.

Artem K. Grebenko*, Konstantin A. Motovilov, Anton V. Bubis, Albert G. Nasibulin**

**Gentle patterning approaches towards environmental friendliness and enhanced compatibility**

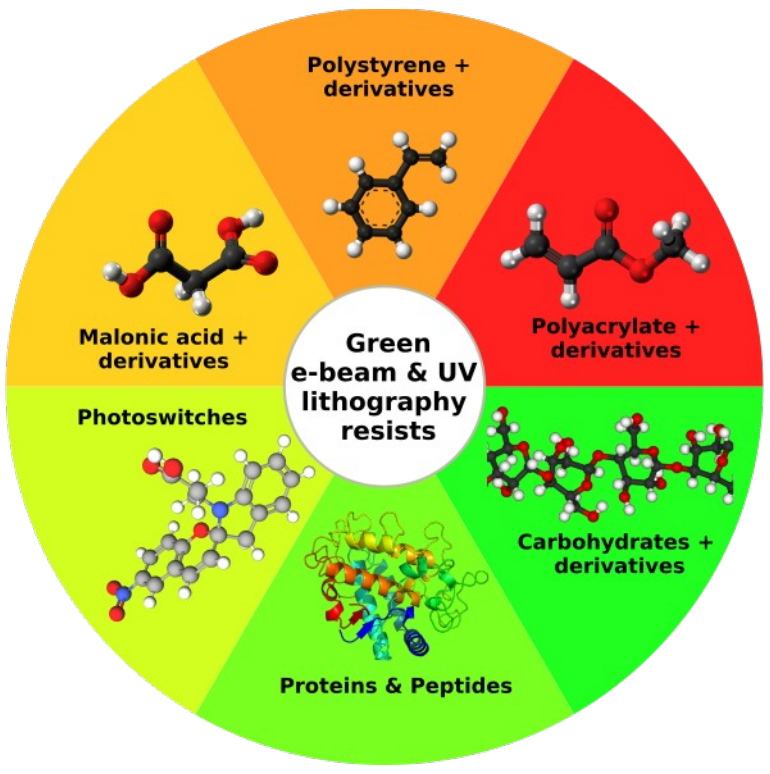